\documentclass[sigconf]{acmart}
\usepackage{subfigure}
\usepackage{threeparttable}
\usepackage{multirow}
\usepackage{multicol}

\AtBeginDocument{%
  \providecommand\BibTeX{{%
    \normalfont B\kern-0.5em{\scshape i\kern-0.25em b}\kern-0.8em\TeX}}}

\copyrightyear{2021}
\acmYear{2021}
\setcopyright{acmcopyright}\acmConference[MM '21]{Proceedings of the 29th ACM
International Conference on Multimedia}{October 20--24, 2021}{Virtual Event, China}
\acmBooktitle{Proceedings of the 29th ACM International Conference on Multimedia
(MM '21), October 20--24, 2021, Virtual Event, China}
\acmPrice{15.00}
\acmDOI{10.1145/3474085.3475558}
\acmISBN{978-1-4503-8651-7/21/10}
\begin{document}
\fancyhead{}
%%
%% The "title" command has an optional parameter,
%% allowing the author to define a "short title" to be used in page headers.
\title[Cross Modal Compression et al.]{Cross Modal Compression: Towards Human-comprehensible Semantic Compression}

%%
%% The "author" command and its associated commands are used to define
%% the authors and their affiliations.
%% Of note is the shared affiliation of the first two authors, and the
%% "authornote" and "authornotemark" commands
%% used to denote shared contribution to the research.
\author{Jiguo Li}
% \authornote{Both authors contributed equally to this research.}
\email{jiguo.li@vipl.ict.ac.cn}
% \orcid{0002-1447-4798}
\affiliation{%
  \institution{Institute of Computing Technology, Chinese Academy of Sciences}
  \city{Beijing}
  \country{China}
}
\additionalaffiliation{
\institution{University of Chinese Academy of Sciences, Beijing 100049, China}
  \city{Beijing}
  \country{China}
}

\author{Chuanmin Jia}
\authornote{Chuanmin Jia is the corresponding author.}
\email{cmjia@pku.edu.cn}
\affiliation{%
 % \department{the Institute of Digital Media, School of Electronic Engineering and Computer Science}
  \institution{Peking University}
  \city{Beijing}
  \country{China}
}

\author{Xinfeng Zhang}
\email{xfzheng@ucas.ac.cn}
\affiliation{%
% \department{the School of Computer Science and Technology}
  \institution{University of Chinese Academy of Sciences}
  \city{Beijing}
  \country{China}
}

\author{Siwei Ma}
\email{swma@pku.edu.cn}
\affiliation{%
% \department{the Institute of Digital Media, School of Electronic Engineering and Computer Science}
  \institution{Peking University}
  \city{Beijing}
  \country{China}
}
\additionalaffiliation{%
  \institution{Information Technology R\&D Innovation Center of Peking University, Shaoxing 312000, China}
  \city{Shaoxing 312000}
  \country{China}
}

\author{Wen Gao}
\email{wgao@pku.edu.cn}
\affiliation{%
% \department{the Institute of Digital Media, School of Electronic Engineering and Computer Science}
  \institution{Peking University}
  \city{Beijing}
  \country{China}
}

%%
%% By default, the full list of authors will be used in the page
%% headers. Often, this list is too long, and will overlap
%% other information printed in the page headers. This command allows
%% the author to define a more concise list
%% of authors' names for this purpose.
\renewcommand{\shortauthors}{Li, et al.}

%%
%% The abstract is a short summary of the work to be presented in the
%% article.
\begin{abstract}
    Traditional image/video compression aims to reduce the transmission/storage cost with signal fidelity as high as possible. However, with the increasing demand for machine analysis and semantic monitoring in recent years, semantic fidelity rather than signal fidelity is becoming another emerging concern in image/video compression. 
    With the recent advances in cross modal translation and generation, in this paper, we propose the cross modal compression~(CMC), a semantic compression framework for visual data, to transform the high redundant visual data~(such as image, video, etc.) into a compact, human-comprehensible domain~(such as text, sketch, semantic map, attributions, etc.), while preserving the semantic. 
    Specifically, we first formulate the CMC problem as a rate-distortion optimization problem. Secondly, we investigate the relationship with the traditional image/video compression and the recent feature compression frameworks, showing the difference between our CMC and these prior frameworks. Then we propose a novel paradigm for CMC to demonstrate its effectiveness. The qualitative and quantitative results show that our proposed CMC can achieve encouraging reconstructed results with an ultrahigh compression ratio, showing better compression performance than the widely used JPEG baseline. 
\end{abstract}

%%
%% The code below is generated by the tool at http://dl.acm.org/ccs.cfm.
%% Please copy and paste the code instead of the example below.
%%
\begin{CCSXML}
<ccs2012>
   <concept>
       <concept_id>10010147.10010257.10010258.10010259.10010264</concept_id>
       <concept_desc>Computing methodologies~Supervised learning by regression</concept_desc>
       <concept_significance>300</concept_significance>
       </concept>
   <concept>
       <concept_id>10002950.10003712.10003713</concept_id>
       <concept_desc>Mathematics of computing~Coding theory</concept_desc>
       <concept_significance>300</concept_significance>
       </concept>
 </ccs2012>
\end{CCSXML}

\ccsdesc[300]{Computing methodologies~Supervised learning by regression}
\ccsdesc[300]{Mathematics of computing~Coding theory}

%%
%% Keywords. The author(s) should pick words that accurately describe
%% the work being presented. Separate the keywords with commas.
\keywords{Image/Video Compression, Deep Neural Networks, Multimedia}

%\begin{teaserfigure}
%  \centering
%  \includegraphics[width=\linewidth]{./fig/CMC_framework_out.pdf}
%  \caption{Illustration of our proposed Cross Modal Compression~(CMC) framework. The compressed representation in the compression domain is a compact, common, and human-comprehensible feature~(such as text, sketch, semantic map, attributions. etc.) which can be losslessly encoded into a bitstream. The whole framework consists of four parts: CMC encoder, CMC decoder, entropy encoder, and entropy decoder.}
% \label{fig:framework}
% \end{teaserfigure}

%% A "teaser" image appears between the author and affiliation
%% information and the body of the document, and typically spans the
%% page.
% \begin{teaserfigure}
%   \includegraphics[width=\textwidth]{sampleteaser}
%   \caption{Seattle Mariners at Spring Training, 2010.}
%   \Description{Enjoying the baseball game from the third-base
%   seats. Ichiro Suzuki preparing to bat.}
%   \label{fig:teaser}
% \end{teaserfigure}

%%
%% This command processes the author and affiliation and title
%% information and builds the first part of the formatted document.
\maketitle

\section{Introduction}\label{sec:introduction}

% Traditional image/video compression aims to reduce the transmission and storage cost with as high as possible fidelity on pixel level. The traditional image/video coding framework is baed on the block splitting, intra/inter prediction, transformation and entropy coding, to encode the raw pixel value into the low-correlational bit stream~\cite{wallace1990overview, marcellin2000overview, sullivan2012overview}.
% The data explosion recently makes that most of the data are only accessed by the machine, not human, especially the image and video data, indicating the semantic rather than the signal fidelity is becoming the most important concern in the compression. 
Data explosion makes more and more data be accessed by machines instead of humans, especially for image and video data. Therefore, the semantic fidelity~\cite{chen2019learning}, rather than signal fidelity, is becoming a more important metric in image/video compression. Because signal fidelity aims to the human visual system and is widely used in traditional compression.
Besides, monitoring the semantic information, such as the identification, human traffic or car traffic, rather than the raw signal, is becoming the main concern of most applications, which is named \textit{semantic monitoring} in this paper.
However, the traditional block-based image/video compression frameworks~\cite{wallace1990overview,marcellin2000overview,sullivan2012overview} mainly optimize the signal fidelity under certain rate constrain, cannot meet the emerging demand of machine analysis and semantic monitoring. 
Recent feature compression frameworks~\cite{redondi2016compress, chen2019lossy} encode the ultimate/intermediate features of deep neural networks into bitstream via quantization and entropy coding, to concentrate on the semantic fidelity for machine analysis. However, feature compression has three limits:
~(1) it is mostly task-specific so the feature is difficult for multi-task analysis;~(2) it is not human-comprehensible, so further analysis is necessary for semantic monitoring;~(3) the evidence is not enough for these features to reconstruct the data on the semantic level. Therefore in this paper, we propose the cross modal compression framework for human-comprehensible semantic compression to conquer these limits.

%\begin{figure}[t]
%  \centering
%  \includegraphics[width=\linewidth]{./fig/three_scenarios.png}
%  \caption{Three common usage scenarios: (row 1st) machine analysis; (row 2nd) human semantic monitoring; (row 3rd) human raw signal monitoring.}
%  \label{fig:scenarios}
%\end{figure}

The traditional image/video compression framework is a block-based hybrid architecture, including the following submodules: block splitting, prediction, transformation, quantization, entropy encoding~\cite{wallace1990overview,marcellin2000overview,sullivan2012overview}. It simultaneously optimizes the pixel level fidelity with the metric of peak signal to noise ratio~(PSNR), and the transmission or storage cost~(bitrate). The traditional compression framework assumes that we need to reconstruct the original signal from the compression bitstream every time it is accessed, no matter for what we access the data. However, in the machine analysis, such as the retrieval for large scale surveillance video, reconstruction may be unnecessary if we can analyze the data in the compression domain. But the compression domain in traditional frameworks is the bitstream, which cannot be analyzed easily.

Feature compression is proposed to compress the semantic features so that we can analyze the data in the compression domain, without the need to reconstruct the signals. 
The ultimate feature compression~\cite{redondi2016compress} compress the task-specific ultimate features into a bitstream, then these features are stored or transmitted for the future intelligent analysis. 
The raw data reconstruction is unnecessary because the task-specific semantic representation is accessible for the following analysis tasks.
However, the features here are mostly task-specific, and a new feature is needed if a new task is added in our intelligent analysis, as illustrated in Table~\ref{tab:comparison}.
To overcome this limit, the intermediate feature compression~\cite{chen2019lossy} is proposed to extract the intermediate features from the intermediate layers of the deep model, rather than the ultimate layer, making the features more common for multi-tasks analysis.
However, the features here are not human-comprehensible, so further processing is needed for semantic monitoring. Besides, the evidence for reconstructing the raw data from these intermediate features in the semantic level is not enough.

Motivated by the new demands for machine analysis and semantic monitoring, we propose the cross modal compression~(CMC) to compress the high redundant data~(such as images, videos. etc.) into a compact, common, and human-comprehensible compression domain~(such as text, sketch, semantic map, attributions. etc.). With this compression domain, CMC has these advantages:~(1) we can compress the raw data with ultra-high compression ratio while preserving the semantic;~(2) this common representation can be used for multiple machine analysis tasks;~(3) our compact representation is human-comprehensible, so it can be used for semantic monitoring without further processing;~(4) image/video reconstructions, especially image reconstruction from text/sketch/attributions, video reconstruction from semantic maps, have been well-studied, providing enough evidence to reconstruct raw data from the human-comprehensible compression domains.

In general, our contributions in this paper can be summarized as follows:
\begin{enumerate}
  \item We propose a new framework, cross modal compression~(CMC), for human-comprehensible semantic compression to meet the emerging demands and formulate the semantic compression as a rate-distortion optimization problem.
  \item We propose a novel paradigm for cross modal compression by compressing the images into the text because the text representation is compact, common, and human-comprehensible. Recent works about text-to-image generation also provide evidence to reconstruct images from the text on the semantic level.
  \item Qualitative and quantitative results demonstrate the effectiveness of our proposed CMC, showing better compression performance than the widely used JPEG baseline.
\end{enumerate}
% The rest of this paper is organized as follows: Section~\ref{sec:related_works} reviews the related works, including traditional image/video compression, feature compression frameworks and cross modal translation; Section~\ref{sec:cmc} formulates the problem and differentiates our problem with the related ones; Section~\ref{sec:image_text_image} presents a novel paradigm by compressing the images into text domain; Section~\ref{sec:experimental_results} shows the qualitative and quantitative to demonstrate the effectiveness of our proposed method; Section~\ref{sec:conclusion} concludes the paper and discusses the future directions about our work.

\begin{figure*}[t]
  \centering
  \includegraphics[width=\linewidth]{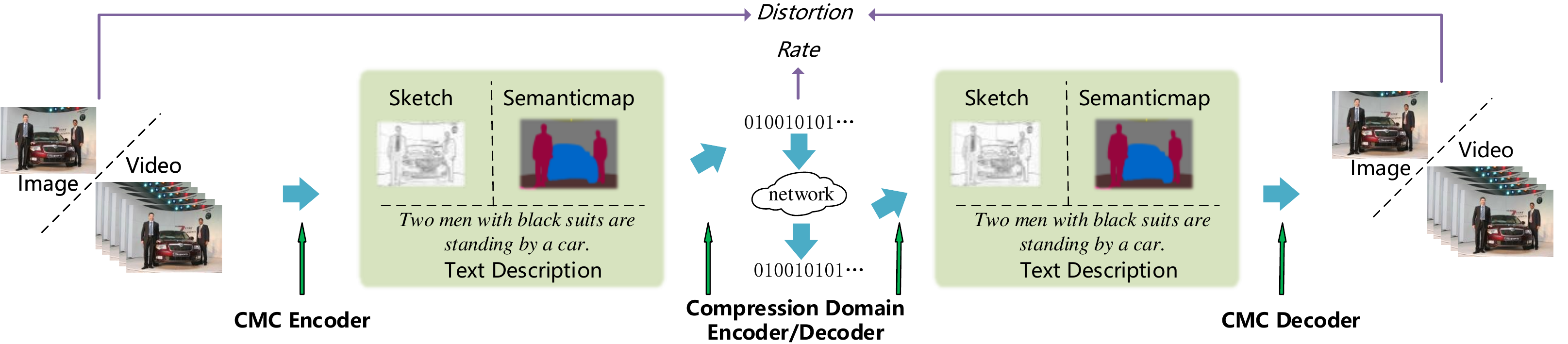}
  \caption{Illustration of our proposed Cross Modal Compression~(CMC) framework. The compressed representation in the compression domain is a compact, common, and human-comprehensible feature~(such as text, sketch, semantic map, attributions. etc.) which can be losslessly encoded into a bitstream. The whole framework consists of four parts: CMC encoder, CMC decoder, entropy encoder, and entropy decoder.}
  \label{fig:framework}
\end{figure*}

\section{Related Works}\label{sec:related_works}
\subsection{Traditional Image/Video Compression}
Traditional image/video compression technologies have been widely applied in our daily life for several decades, which has brought up a series of industry standards, such as JPEG~\cite{wallace1990overview}, JPEG2000~\cite{marcellin2000overview}, TPG~\cite{yuan2018dilated} for image and AVC~\cite{wiegand2003overview}, HEVC~\cite{sullivan2012overview}, AVS2~\cite{gao2014overview} for video. 
% Here we mainly review JPEG because in this paper we focus most of the attention on the image compression. 
% JPEG is an image compression standard for general purposes, including variable resolutions, various color spaces, and different transmission bandwidths, taking the efficiency of both the software and hardware into consideration. It consists of well-known technologies, including $8\times 8$ discrete cosine transformation~(DCT), quantization, and Huffman coding. Following the JPEG standard, several image compression standards with a high compression ratio are specified due to the emerging demand in multimedia applications. JPEG2000~\cite{marcellin2000overview} provides higher compression performance, and also allows extraction of different resolutions as well as the editing or processing applications. 
JPEG and JPEG2000 are image compression standards for general purposes, including variable resolutions, various color spaces, and different transmission bandwidths, taking the efficiency of both the software and hardware into consideration. It consists of well-known technologies, including $8\times 8$ discrete cosine transformation~(DCT)~\cite{ahmed1974discrete}, quantization, and Huffman coding.
In addition to the image compression standards, several video compression standards have also been set, including the widely applied H.264~\cite{wiegand2003overview}, AVS2~\cite{gao2014overview} and H.265~\cite{sullivan2012overview}. They are all block-based hybrid coding frameworks, in which quad-split, intra/inter prediction, DCT, quantization, entropy coding~\cite{huffman1952method} are used to reduce the statistical redundancy, spatial/temporal redundancy, and perceptual redundancy.
Recently, deep-based frameworks have also been proposed to compress the images via end-to-end optimization~\cite{ma2019image, balle2017end, balle2018variational}, and these methods have shown the potential that surpasses widely used state-of-the-art image codecs, such JPEG, JPEG2000 and HEVC intra, although they have the difficulties to be standardized due to the non-uniqueness of the model's parameters and structures. Also, the submodules in video compression frameworks are investigated by embedding the DNNs in the them, including intra prediction~\cite{li2018fully}, inter prediction~\cite{zhao2018enhanced}, loop filter~\cite{jia2019content}, quantization~\cite{alam2015perceptual} and entropy encoding~\cite{puri2017cnn}. 
% In this paper, we mainly compare our proposed methods with the traditional standardized compression methods.
\subsection{Feature Compression}
% With the explosive growth of cloud-based intelligent analysis applications, such as smart city and surveillance analysis, feature compression paradigms are proposed to convey the deep semantic features rather than the traditional compressed bitstream on the fact that most of the analysis applications rely on the deep semantic features. 
% Redondi~\cite{redondi2016compress} compared the two different paradigms for the visual sensor networks: (1) the compress-then-analyze paradigm which conveys the compressed bitstream from the front-end devices to the back-end servers and analyzes the data from the scratch in the back-end; (2) the analyze-then-compress paradigm which processes the data in the front-end devices and conveys the extracted features to the back-end servers. This proposed paradigm can reduce the network load and allocate the computation into the front-end nodes, but this work only investigated the hand-crafted feature, such as SIFT~\cite{lowe2004distinctive}. 
To alleviate the network load and the computation in the back-end, Redondi~\textit{et al.}~\cite{redondi2016compress} proposed an analyze-then-compress paradigm instead of the traditional compress-then-analyze framework.
But this work only investigated the hand-crafted feature, such as SIFT~\cite{lowe2004distinctive}. 
% Choi~\textit{et al.}~\cite{choi2018deep} examined to lossy compress the deep ultimate features for the object detection.
Choi~\textit{et al.}~\cite{choi2018deep} examined to lossy compress the deep ultimate features for the object detection. However, the ultimate feature is usually task-specific, making it difficult to process increasing intelligent analysis tasks. To conquer this problem, Chen~\textit{et al.}~\cite{chen2019lossy} presented to transmit the intermediate features, which are derived from the intermediate activation layers of the DNNs, rather than the ultimate layer to enable a good balance among the transmission load, computing load, and the generalization ability for different intelligent analysis tasks. 
The features from the lower layers of DNNs are less abstract and less task-specific because a DNN can be viewed as a cascaded feature extractor~\cite{chen2019lossy}. Therefore the intermediate feature can be adopted by various tasks and is with better generalization ability than the ultimate feature. 
However, a deficiency of the intermediate features is that it cannot be understood by the human, so the raw data are needed when the semantic monitoring is necessary for the human.

\subsection{Cross Modal Translation}
Cross modal translation aims to convert the data from one modality to another with semantic consistency, which is an emerging topic in recent years, including image-to-text~\cite{hossain2019comprehensive, vinyals2015show, jia2015guiding, wang2016image, ren2017deep}, text-to-image~\cite{reed2016generative,zhang2018stackgan++,xu2018attngan}, video-to-text~\cite{wang2018show, pan2016jointly}, text-to-video~\cite{pan2017create}, sketch-to-image~\cite{liu2018auto}, image-to-sketch~\cite{wang2017bayesian}, etc. In this paper we mainly review the works on image-to-text and text-to-image translation. 
\textit{Image-to-text} translation~(I2T), also known as image caption, represents the images as a syntactically and semantically correct sentence, which is a compact, human-comprehensible form. Before DNNs were adopted in I2T, template-based methods~\cite{farhadi2010every} and retrieval-based methods~\cite{gong2014improving} are two main approaches for I2T~\cite{hossain2019comprehensive}. Once deep based methods were applied in I2T, it showed incomparable performance via an encoder-decoder framework~\cite{vinyals2015show, sutskever2014sequence}.
% CNN-based image encoder and an RNN-based language decoder~\cite{vinyals2015show}. Similar to the framework of the neural machine translation~\cite{sutskever2014sequence}, Vinyals~\textit{et al.}~\cite{vinyals2015show} proposed an encoder-decoder framework to use a pretrained CNN model with batch normalization~\cite{ioffe2015batch} to extract the image representation and an LSTM-based sentence generator to decode the caption word by word. Following this work, a series of works were introduced to improve the language decoder~\cite{jia2015guiding, wang2016image}. 
% Recently, Ren~\textit{et al.}~\cite{ren2017deep} introduced the reinforcement learning into the I2T framework by modeling the language generation as a decision-making problem and training a ``policy network'' and a ``value network'' via a novel visual-semantic embedding reward. 
\textit{Text-to-Image} translation~(T2I), also known as image generation from the text, aims to synthesize fine-grained images from the text descriptions with semantic consistency. Reed~\textit{et al.}~\cite{reed2016generative} demonstrated it feasible to synthesize images with semantic consistency from the text consistency via generative adversarial networks~(GANs)~\cite{goodfellow2014generative}, although the resolution of generated images is only $64\times 64$.  Based on the advances in image generation~\cite{salimans2016improved, arjovsky2017wasserstein}, the following T2I works~\cite{zhang2017stackgan, zhang2018stackgan++, xu2018attngan} succeeded to synthesize images with higher resolution and more details based on a progressive framework.
% Zhang~\textit{et al.}~\cite{zhang2017stackgan} synthesized the images with a resolution of $256\times 256$ from the text descriptions by generating high-resolution images with two increasing stages and discriminating the generated data distribution in both conditional and unconditional settings. Following this promising idea that splitting the generation process of a high-resolution image into several super resolution submodules, Zhang~\textit{et al.}~\cite{zhang2018stackgan++} synthesized photo-realistic images from the text descriptions with a three-stage generator and three discriminators for increasing resolutions, achieving superior performance to the previous works. Subsequently, Xu~\textit{et al.}~\cite{xu2018attngan} introduced the attention mechanism into T2I and synthesized fine-grained images from the text descriptions. 
%\indent The remarkable advance of cross modal translation forms the basis of our cross modal compression.
 % and we take much inspiration from these prior works. 
% The further advances of cross modal translation will also help boost the performance and practicability of our proposed CMC.

\section{Cross Model Compression~(CMC)}\label{sec:cmc}
\subsection{Problem Formulation}
% Cross modal compression~(CMC) aims to compress the high redundant data into a compact, common, human-comprehensible representation, which can be adopted by various machine analysis applications as well as the semantic monitoring. 
Data compression aims to reduce the transmission or storage cost with certain fidelity, which can be formulated as follows:
\begin{equation}
  g = D + \lambda R,
\end{equation}
where $R$ is the bitrate, $D$ denotes the distortion, which is evaluated in pixel level in traditional image/video compression. 
Cross modal compression~(CMC) aims to compress the high redundant data into compact, common, human-comprehensible representation, which can be adopted by various machine analysis applications. 
In CMC, a compression domain $\mathbb{Y}$, where the representation is compact, common, and human-comprehensible, is defined. In this domain, the compressed representation can be losslessly encoded as a bitstream. As illustrated in Fig.~\ref{fig:framework}, the framework consists of four submodules: CMC encoder, CMC decoder, entropy encoder, and entropy decoder. CMC encoder compresses the raw signal into a compact and human-comprehensible representation, which can be decoded by the CMC decoder to reconstruct the signal with semantic consistency.
% The bitrate of the compressed representation is the entropy of the bitstream and the distortion is the error between the raw signal and the reconstructed signal.
The bitrate is optimized by finding a compact compression domain, while the distortion is optimized by preserving the semantic in CMC encoder and decoder.

\begin{table*}
  \caption{Comparison with related compression frameworks}
\begin{center}
  \begin{threeparttable}
    \begin{tabular}{c|c|c|c|c|c|c}
      \hline
      \multirow{2}{*}{Methods} & Compression & Multi-task & Human & Frondend & Backend & Data \\
       & Ratio & Analysis & Comprehensible & Load & Load & Reconstruction\\
       \hline
       \hline
       Traditional Compression$^\star$ & Middle & $\checkmark$ & $\times$ & Middle & High &$\checkmark$ \\
       \hline
       Ultimate Feature Compression & High & $\times$ & $\times$ & High & Low & - \\
       \hline
       Intermediate Feature Compression & High & $\checkmark$ & $\times$ & Middle & Middle & - \\
       \hline
       Cross Modal Compression & High & $\checkmark$ & $\checkmark$ & Middle & Middle & $\checkmark$ \\
       \hline
    \end{tabular}
    \begin{tablenotes}
      \item $^\star$ Such as JPEG~\cite{wallace1990overview}, H.264~\cite{wiegand2003overview}, HEVC~\cite{sullivan2012overview}, etc. 
      \item - The evidence for this task is not enough.
    \end{tablenotes}
  \end{threeparttable}
\end{center}
\label{tab:comparison}
\end{table*}

\subsection{Comparison with Related Frameworks}
%\begin{figure}[t]
%  \centering
%  \includegraphics[width=\linewidth]{./fig/cmc_comparison.png}
%  \caption{Comparison of the four transmission frameworks: (a) %traditional compression; (b) ultimate feature compression; (c) %intermediate feature compresson; (d) cross modal compression.}
%  \label{fig:comparison}
%\end{figure}
In this section, we compare our proposed CMC with the related frameworks and show the difference with these prior works, as shown in Table~\ref{tab:comparison}.
% four compression frameworks are illustrated to explain the difference intuitively.
\begin{enumerate}
  \item \textit{Traditional Signal Compression.} 
  % Traditionally, images/videos are acquired by the front-end devices~(such as surveillance cameras, mobile phones. etc.) and lossy compressed by the standardized image/video codecs into bitstream so that the storage and bandwidth can be saved as much as possible. 
  Traditional codecs optimize the pixel level fidelity for human visual perception by minimizing the pixel level metrics. When more and more intelligent analysis applications come to our traditional signal compression systems, the data must be reconstructed so that we can process the images/videos~(recognized, detected, or enhanced) to extract the semantic information. 
  However, the explosive growth of images/videos has made that most of the visual data are not watched by the human but by the intelligent analysis applications. So pixel level optimization may waste the storage and bandwidth due to the information redundant for these intelligent applications.
  \item \textit{Ultimate Feature Compression.} Due to the inefficiency of traditional signal compression in some scenarios where only the analysis results are needed rather than the pixel level representation, ultimate feature compression~\cite{redondi2016compress} were proposed based on the tenet that most of the visual analysis tasks can be carried with the ultimate feature. In this \textit{Analyse-then-Compress} paradigm, ultimate features are extracted on the front-end devices and then delivered to the back-end server to enable the intelligent analysis tasks. 
  %Consequently, the computational intensive feature extraction is deployed on the distributed edge nodes and the load of cloud center~(usually the bottleneck for the visual systems) can be alleviated dramatically. Nevertheless, the reconstruction from the compressed bitstream is entailed if human raw signal monitoring is needed.  
  Although this paradigm can alleviate the load of the cloud center, this framework may suffer from an obstacle that the ultimate features are usually task-specific so it is difficult to adopt them in the tasks except for the specific one. Moreover, deploying various deep models in front-end devices will make the systems bloated.
  \item \textit{Intermediate Feature Compression.} 
  %To apply the transmitted features in various tasks, Chen~\textit{et al.}~\cite{chen2019lossy} proposed to compress and convey the intermediate features instead of the ultimate ones. 
  The intermediate features, extracted from the intermediate layers of the DNNs, are less abstract and can be applied in various tasks. By conveying the intermediate features, the computational load on front-end devices and back-end cloud servers can be well balanced, indicating the flexibility to deploy this kind of framework. 
  Besides, lossy compression for the intermediate features is also conducted to encode the feature into bitstream to be stored or transmitted.
  However, the intermediate features cannot be understood by humans so they cannot be used for human semantic monitoring directly, and further processing and analysis are necessary. 
  % although humans only care for the semantic. 
  \item \textit{Cross Modal Compression.} To accomplish a compact, common, and human-comprehensible representation for image/video data, we propose the cross modal compression~(CMC), as illustrated in Fig.~\ref{fig:framework}. The image/video $x, x\in\mathbb{X}$, is firstly transformed into $y$, $y\in\mathbb{Y}$. $y$ is a compact and human-comprehensible representation, such as text, sketch, semantic map, or attributions,  which can be adopted in semantic monitoring directly. Besides, $y$ is a common feature and can be adopted for various intelligent analysis applications. When transmission, $y$ can be compressed further into a bitstream by an entropy encoder. If the raw images/videos are needed, we can also reconstruct the raw data from $y$ with semantic consistency. Our proposed CMC is different from all the above previous frameworks, as illustrated in Table~\ref{tab:comparison}.
\end{enumerate}
CMC is a novel framework for visual data compression, such as image and video, which is a human-comprehensible semantic compression framework. We can design the specific submodules for CMC encoder/decoder, entropy encoder/decoder, when the source data domain, compression domain are determined. 
In the following, we will introduce a paradigm to compress the images into the text domain, which is compact, common, and human-comprehensible. 
% We also attempt to reconstruct the raw image data and evaluate the reconstruction performance.
%\begin{table}
%  \begin{tabular}[c|c|c|c|c]
%    \hline
%    
%  \end{tabular}
%\end{table}

 \begin{figure*}[t]
  \centering
    \includegraphics[width=\linewidth]{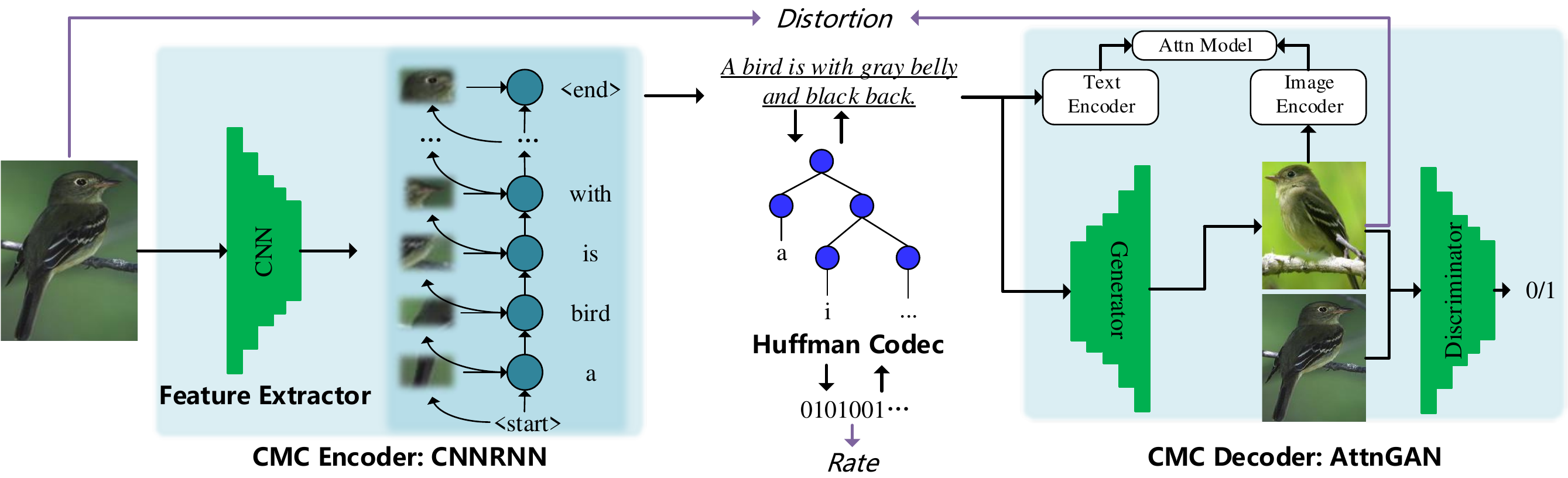}
    \caption{Illustration for a paradigm of CMC: Image-Text-Image~(ITI)}
    \label{fig:illustration_iti}
  \end{figure*}
  
\section{Image-Text-Image: a paradigm of CMC}\label{sec:image_text_image}
In this section, we introduce a paradigm for our proposed CMC. With the advances of image-to-text translation, understanding an image and describing it with natural languages is possible. Meanwhile,
prior works on text-to-image translation have shown enough evidence for reconstructing the image from the text on the semantic level.
We define the compression domain $\mathbb{Y}$ as the text domain in our CMC framework and propose the Image-Text-Image~(ITI) compression framework for cross modal image compression. As illustrated in Fig.~\ref{fig:framework}, there are mainly three submodules in the framework:~(1)~CMC encoder to compress the raw image into a text description;~(2)~Lossless compression in compression domain to encode the text description into a bitstream and decode the bitstream into language;~(3)~CMC decoder to reconstruct images from the text description. 
\subsection{CMC Encoder}
CMC encoder aims to compress the data from the image/video domain into a compact domain, which is the text domain in this paradigm.
With the advance of image caption in recent years, a CNN-RNN with attention model referring to~\cite{vinyals2015show} is used in our CMC encoder, as shown in Fig.~\ref{fig:illustration_iti}. Given an input image $v_i, i=1,2,3,\ldots,N$, $N$ is the number of the samples, a downsampled feature $f_i$ can be obtained by a CNN based feature extractor:
\begin{equation}
  f_i = \text{CNN}(v_i),
\end{equation}
where the feature $f_i=\{f_{i,1}, f_{i,2}, f_{i,3}, \ldots,f_{i,L}\}$ contains $L$ vectors that correspond different positions on the image. Following the CNN feature extractor, an RNN with attention is leveraged to decode the predicted words step by step. As introduced in~\cite{Bahdanau2015neural} and~\cite{vinyals2015show}, at time step $s_t$, the attention mechanism can be formulated as follows (described as ``soft attention'' in~\cite{vinyals2015show}):
\begin{equation}
  \mathbb{E}_{p(s_t|a)}[\hat{z}_t] = \sum_{j=0}^L \alpha_{t,j}a_j,
\end{equation}
where $\hat{z}_t$ is the context vector at time step $s_t$, dynamically representing the relevant part of the image, $a_j, j=1,2,3,\ldots, L$, denotes the annotation vector that corresponds to the extracted feature $f_{i,j}$ at different image locations. $\alpha_{t,j}$ is a weight for each annotation vector $a_j$, which can be computed by an attention model $f_{attn}$:
\begin{align}
 e_{t,j} &= f_{att}(a_j, h_{t-1}) \\
 \alpha_{t, j} &= \frac{\exp{(e_{t,j})}}{\sum_{k=1}^L\exp{(e_{t,k})}},
\end{align}
where $h_{t}$ is the hidden output of the RNN at time step $t$, $e_{t,j}$ is the attention output before the softmax layer.

 At the first time step, the starting word ``$<$start$>$'' is feed into the RNN to compute the attention map, build the context and decode the following words. The RNN output one word at each step until the ending word ``$<$end$>$'' is outputted.
% \begin{figure}[t]
%   \centering
%   \includegraphics[width=\linewidth]{./fig/imagetext.png}
%   \caption{CMC Encoder: CNNRNN with Atttention~\cite{vinyals2015show}}
%   \label{fig:imagetext}
% \end{figure}
\subsection{Lossless Compression in Compression Domain}
We think that the compression domain is more semantically compact, compared with the source data domain. Conversion from the source data domain to the compression domain reduces the semantic redundancy. However, there is still statistical redundancy in the compression domain. According to Shannon's information theory~\cite{shannon1948mathematical}, the optimal code length for a symbol is $\log_2 P$, where $P$ is the probability of the symbol. In our compression domain, the distribution of the text characters is a prior probability, which can be counted from the training set. 
% Huffman coding~\cite{huffman1952method} can be used to reduce the statistical redundancy, which can be described as follows:
% \indent \textit{If the probability of symbol $s_m$ is $P_m$ and its code length is $L_m$, we have $\sum_{m}^M P_m = 1$. The average code length $L_{avg}=\sum_m^M P_m L_m$. A minimum redundancy codec, which minimizes $L_{avg}$ can be constructed by sorting the symbols in descending probability, sorting the codewords in ascending length, and mapping them two one by one.}
In our paradigm, Huffman coding~\cite{huffman1952method} can be used to reduce the statistical redundancy. The statistical probability of the symbols can be obtained from the training set under the assumption that the training set and the testing set have the same distribution. 
With the symbol probability, we can construct the Huffman tree and design the Huffman encoder based on the Huffman tree.
% The symbol is the character or the word. Optimum binary codec is constructed to reduce the statistical redundancy.
 The paired Huffman decoder uses the same Huffman tree with the encoder's. It is worth mentioning that Huffman coding is lossless, so we can reconstruct the text without any information loss.

\subsection{CMC Decoder}
% \begin{figure}[t]
%   \centering
%   \includegraphics[width=\linewidth]{./fig/textimage.png}
%   \caption{CMC Decoder: AttnGAN~\cite{xu2018attngan}}
%   \label{fig:textimage}
% \end{figure}
CMC decoder aims to reconstruct the data from the compression domain. In our paradigm, we need to reconstruct the image from the text description with semantic consistency. With the recent advances in text-to-image generation~\cite{zhang2017stackgan, zhang2018stackgan++, xu2018attngan}, we use AttnGAN~\cite{xu2018attngan} in our CMC decoder to reconstruct images from the text due to its promising performance on text-to-image generation. AttnGAN integrates the attention mechanism into the generator by pretraining a text encoder and an image encoder to extract position-sensitive features. In the implementation, we use the pretrained Inception-v3 model~\cite{szegedy2016rethinking}, which is trained on Imagenet~\cite{deng2009imagenet}, as the image encoder and train the text encoder on our own dataset, following~\cite{xu2018attngan}.
% the image position-sensitive features are extracted via pretrained Inception-v3 model~\cite{szegedy2016rethinking}, and the text position-sensitive features are extracted by the text encoder which is pretrained on the training set. 
Given the text/image positive-sensitive feature $e/v$, the attention matching score can be calculated as follows:
\begin{align}
  &c  = \text{softmax}(\gamma_1 e^Tv, \text{dim}=0)v^T \\
  &R(c_j, e_j) = \cos(c_j, e_j),
\end{align}
where $c_j$ is a region-context vector dynamically representing the image's subregion related to the $j^{th}$ word, $\gamma_1$ is a factor to control the attention to its relevant sub-regions, $\cos(x,y)={x^T y}/{\|x\|\|y\|}$ is the cosine similarity. The~\textit{attention-driven image-text matching score} between the image and the text description is defined as:
\begin{align}
  R(v, e) = \log{(\sum_{j=1}^J\exp{(\gamma_2 R(c_j, e_j))})^{\frac{1}{\gamma_2}}},
\end{align}
where $J$ is the word number of the text description, $\gamma_2$ is a factor to control the importance of different word-to-subregion pairs.  
\\
\indent
 To synthesize images with a resolution $256\times256$, AttnGAN firstly generates images with a resolution $64\times64$, then upsamples the generated images with ratio 2 and adds the details, until the image with resolution $256\times256$ is generated. 
 % Generating images with increasing resolutions have been demonstrated a promising method to synthesize high resolution realistic images~\cite{zhang2017stackgan, zhang2018stackgan++}.
 It has been demonstrated that generating images with increasing resolutions is a promising method to synthesize high-resolution realistic images~\cite{zhang2017stackgan, zhang2018stackgan++}.
  More details about AttnGAN can be found in~\cite{xu2018attngan}.

  \begin{figure*}[t]
    \centering
      \includegraphics[width=\linewidth]{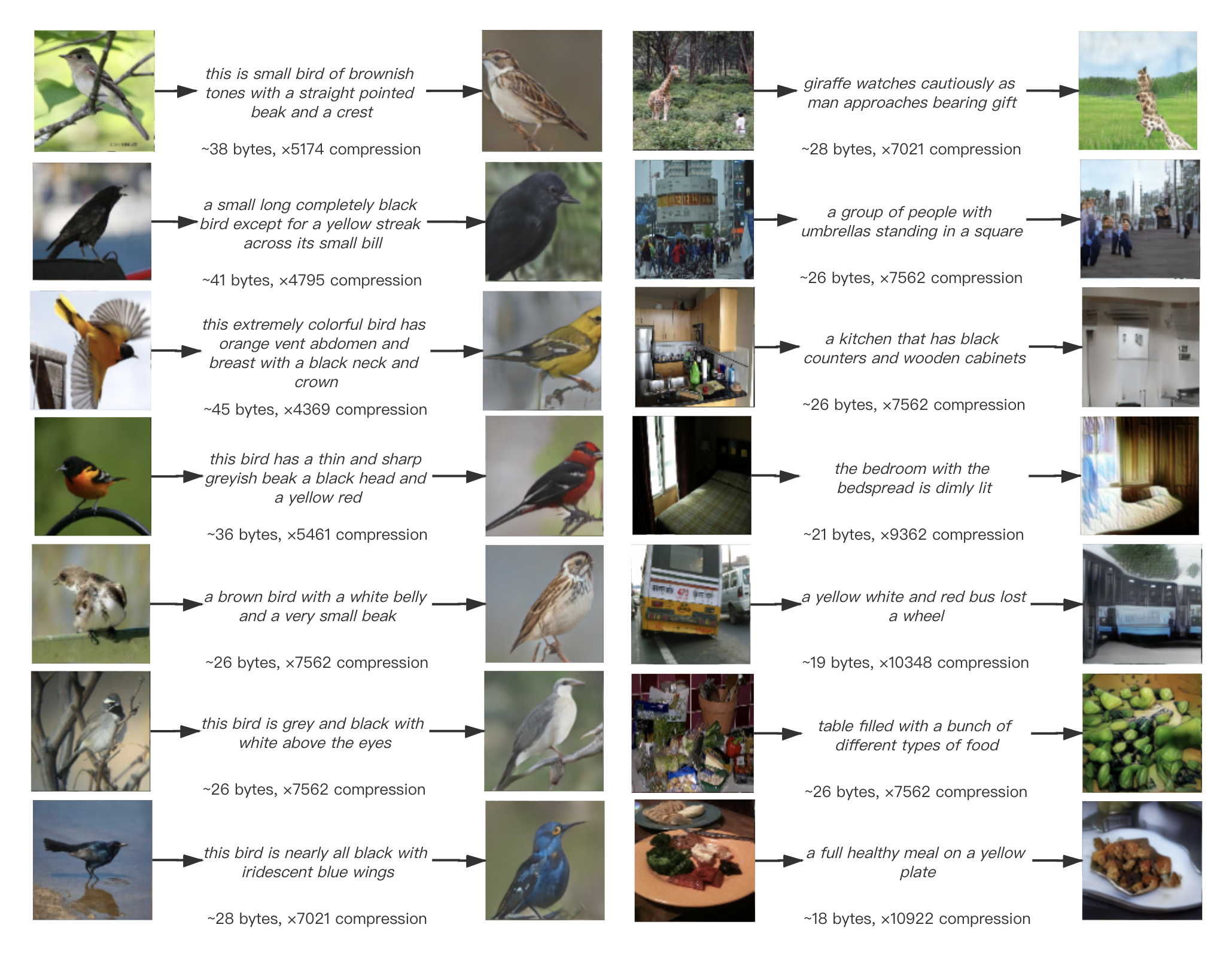}
      \caption{Qualitative results of our ITI framework on CUB-200-2011~(left) and MS COCO~(right). For each sample, we show the raw image, the text representation, and the reconstructed image, subsequently. We also show the \textit{bitrate} and the \textit{compression ratio} under each text.}
      \label{fig:qualitative_results}
  \end{figure*}

 \section{Datasets and Metrics}
 \subsection{Datasets}
 We use MS COCO~\cite{lin2014microsoft} and CUB-200-2011~\cite{WahCUB_200_2011} to evaluate our proposed model's effectiveness.
 MS COCO contains 82783/40504 images for training/testing set, and no less than 5 captions for each image. Images are resized with the resolution of 256$\times$256 for MS COCO. CUB-200-2011 contains 8855/2933 images for training/testing set, and 10 captions for each image. Total 200 classes for CUB-200-2011 are split into 160/40 classes for training/testing. The images are cropped with the annotated bounding box and resized into 256$\times$256 for the following experiments.
 \subsection{Metrics}
We use four metrics to evaluate our proposed method on different levels, which are introduced as follows:
 % inception score~\cite{salimans2016improved} and Fréchet inception distance~\cite{dowson1982frechet, heusel2017gans} to measure the distribution distance between the source data and the reconstructed data, use instance perceptual distance to evaluate the semantic distortion in instance level. Besides, we also report the pixel level reconstruction performance, which is measured by the peak signal-to-noise ratio, to give intuitive comparison results to the JPEG baseline.
 % Rate is also used to measure the compression performance of our framework. Besides, BLEU4 is used to evaluate the performance of the CMC encoder, following~\cite{vinyals2015show}. 
 % These four metrics are introduced as follows.
 \\
 \indent\textit{Peak Signal-to-Noise Ratio~(PSNR)} is defined as:
 \begin{align}
  \text{PSNR} = 10\log_{10}[\frac{(2^{\text{bits}}-1)^2}{\text{MSE}}],
 \end{align}
 where $\text{bits}$ denotes the bit number for a pixel, $\text{MSE}$ denotes mean square error between the source data and the reconstructed data. PSNR is widely used in traditional image/video compression algorithm evaluation.
 \\
 \indent\textit{Inception Score~(IS)}~\cite{salimans2016improved} measures the naturalness and the diversity of the generated images, which is defined as:
 \begin{align}
  \text{IS} &= \exp(\mathbb{E}_\mathbf{x}\mathbb{KL}(p(y|\mathbf{x})|p(y))),
 \end{align}
 where $p(y|\mathbf{x})$ denotes the conditional label distribution, $p(y)$ denotes the marginal distribution, $\mathbf{x}$ denotes the perceptual features.
 \\
 \indent\textit{Fréchet Inception Distance~(FID)}~\cite{heusel2017gans} measure the distribution distance between the source data and the reconstructed data, which is formulated as: 
 \begin{align}
  \text{FID} &= \lvert\lvert{m_1 - m_2}\rvert\rvert^2_2 + \mathrm{Tr}(C_1+C_2-2(C_1C_2)^{\frac{1}{2}}),
 \end{align}
 where $m$/$C$ is the mean/variance of the perceptual features on the testing set. 
 \\
\indent\textit{Instance Perceptual Distance~(IPD)} is used to measure the instance level perceptual distance, because both IS and FID are the set level metrics, neither can evaluate the instance level distortion. IPD is defined as:
\begin{align}
  \text{IPD} & = \mathbb{E}_\mathbf{x}\|\mathbf{x_{re}}-\mathbf{x}\|_2^2,\label{eq:IPD}
\end{align}
where $\mathbf{x}$/$\mathbf{x_{re}}$ denotes the perceptual features for source/reconstructed data. 
 For IS, FID, and IPD in our experiments, we use the pretrained Inception-v3 model which pretrained on Imagenet~\cite{deng2009imagenet} to extract the perceptual features for MS COCO, and use the finetuned Inception-v3 model for CUB-200-2011, following~\cite{xu2018attngan}. In our experiments, we use the implementation of IS and FID in~\cite{xu2018attngan}
 \footnote{IS for MS COCO: https://github.com/hanzhanggit/StackGAN-inception-model, IS for CUB-200-2011: https://github.com/openai/improved-gan, FID for MS COCO and CUB-200-2011: https://github.com/bioinf-jku/TTUR}. We use the same pretrained Inception-v3 model as FID's in the implementation of IPD.

\begin{figure*}[htbp]
  \centering
  \subfigure[Rate-IS$\uparrow$ on MS COCO]{
  \begin{minipage}[t]{0.25\linewidth}
  \centering
  \includegraphics[width=1.6in]{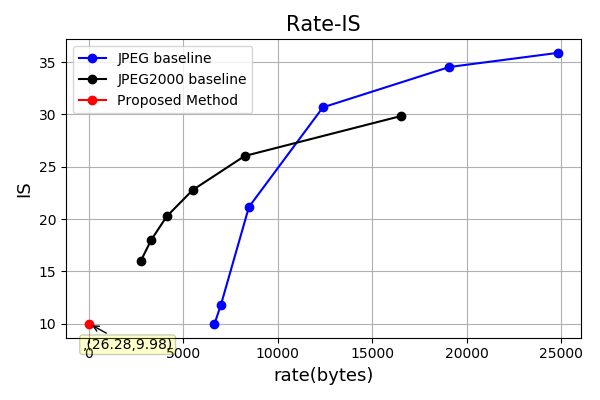}
  %\caption{fig1}
  \end{minipage}%
  }%
  \subfigure[Rate-FID$\downarrow$ on MS COCO]{
  \begin{minipage}[t]{0.25\linewidth}
  \centering
  \includegraphics[width=1.6in]{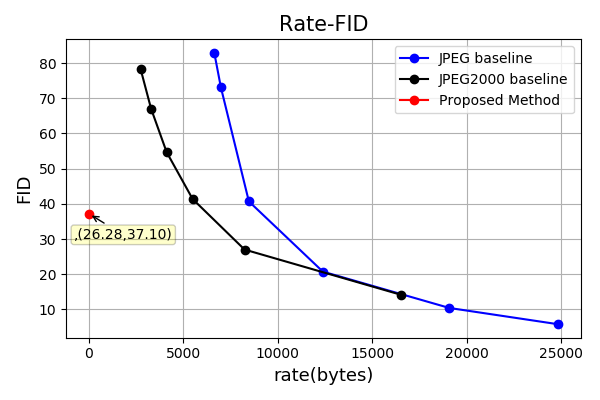}
  %\caption{fig2}
  \end{minipage}%
  }%
  \subfigure[Rate$-$IPD$\downarrow$ on MS COCO]{
  \begin{minipage}[t]{0.25\linewidth}
  \centering
  \includegraphics[width=1.6in]{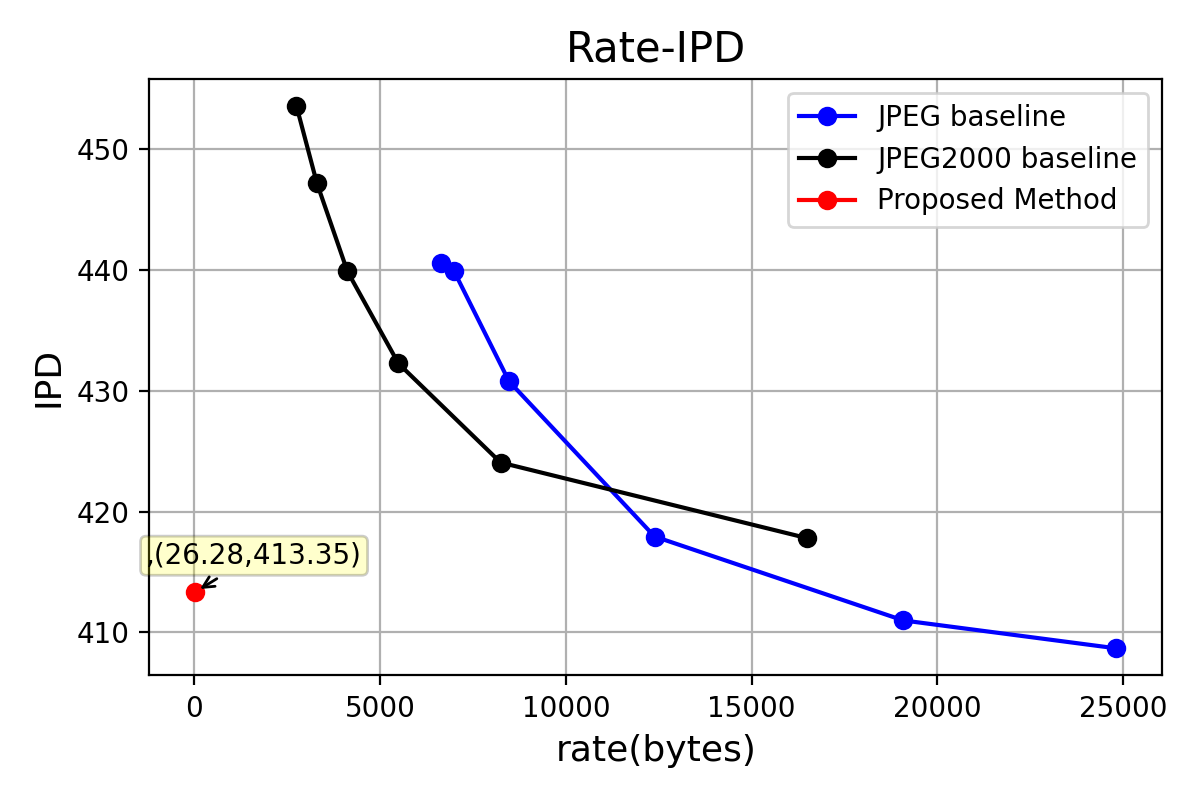}
  %\caption{fig1}
  \end{minipage}%
  }%
  \subfigure[Rate$-$PSNR$\uparrow$ on MS COCO]{
  \begin{minipage}[t]{0.25\linewidth}
  \centering
  \includegraphics[width=1.6in]{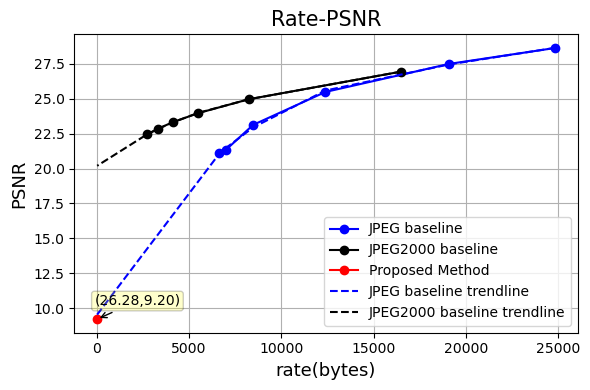}
  %\caption{fig1}
  \end{minipage}%
  }%
  \\               %这个回车键很重要 \quad也可以
  \subfigure[Rate-IS$\uparrow$ on CUB-200-2011]{
  \begin{minipage}[t]{0.25\linewidth}
  \centering
  \includegraphics[width=1.6in]{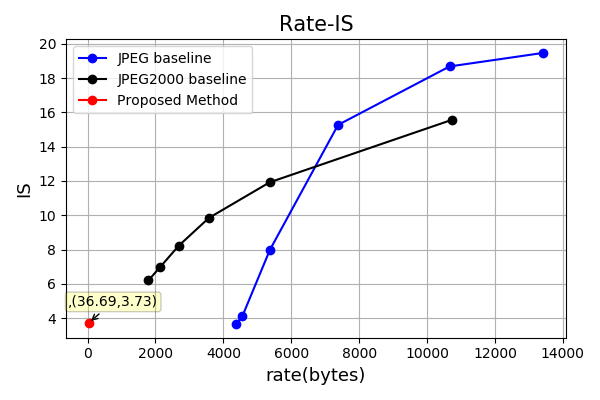}
  %\caption{fig2}
  \end{minipage}
  }%
  \subfigure[Rate-FID$\downarrow$ on CUB-200-2011]{
  \begin{minipage}[t]{0.25\linewidth}
  \centering
  \includegraphics[width=1.6in]{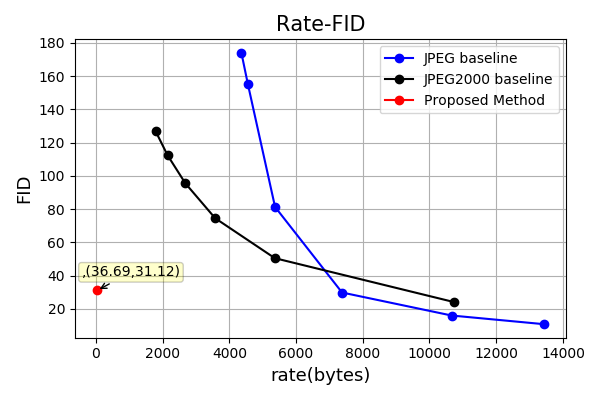}
  %\caption{fig2}
  \end{minipage}
  }%
  \subfigure[Rate-IPD$\downarrow$ on CUB-200-2011]{
  \begin{minipage}[t]{0.25\linewidth}
  \centering
  \includegraphics[width=1.6in]{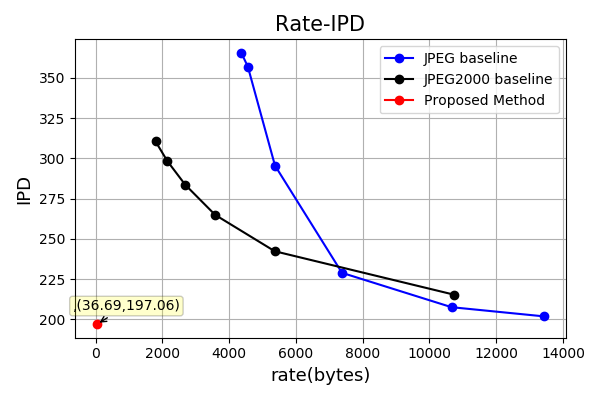}
  %\caption{fig2}
  \end{minipage}
  }%
  \subfigure[Rate-PSNR$\uparrow$ on CUB-200-2011]{
    \begin{minipage}[t]{0.25\linewidth}
    \centering
    \includegraphics[width=1.6in]{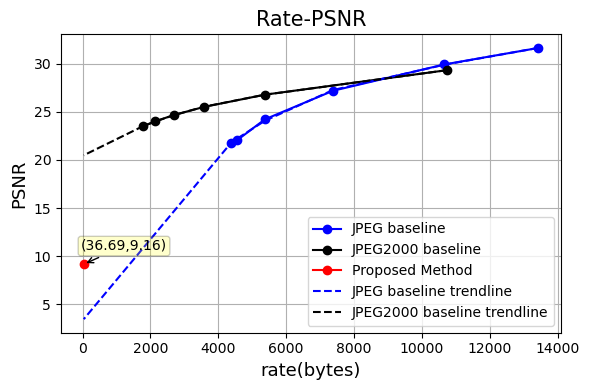}
    %\caption{fig2}
    \end{minipage}
    }%
  \centering
  \caption{Quantitative results: comparison with the JPEG and JPEG2000 baselines~(best view in color).}
  \label{fig:end_to_end}
  \end{figure*}

\section{Experimental Results}\label{sec:experimental_results}
To demonstrate the effectiveness of our proposed paradigm: image-text-image framework for image compression, we conduct experiments to show qualitative and quantitative results on both MS COCO and CUB-200-2011 datasets. 

\subsection{Qualitative Results}
The qualitative results, as well as the compression ratio~\footnote{The compression ratio is calculated under the assumption that the input images are resized to 256$\times$256 with 3 channels.}, on CUB-200-2011 and MS COCO are illustrated in Fig.~\ref{fig:qualitative_results}. Some conclusions can be drawn from the results:

\textit{Our proposed ITI can reconstruct the images well on CUB-200-2011 dataset on instance level with ultrahigh compression ratio.} As shown on the left part of Fig.~\ref{fig:qualitative_results}, most of the reconstructed images are sharp and fine-grained, with a similar appearance as the raw images, although the background may be blurry in the reconstructed ones. Meanwhile, most of the text descriptions are semantically consistent with the raw images and the compression ratio is as high as 4000-7000 times.

\textit{Our proposed ITI can partially reconstruct the images on MS COCO dataset on instance level with ultrahigh compression ratio.} On MS COCO dataset, our ITI can only partially reconstructed the images on the instance level, 
does not show as good performance as that on CUB-200-2011. This is because MS COCO is a more diverse dataset and CUB-200-2011 is a class-specific data for birds. However, as shown on the right of Fig.~\ref{fig:qualitative_results}, our ITI can still partially reconstructed the scene and some key objects in the scene, although the objects may be incomplete and blurry.

Overall, the qualitative results show promising performance on class-specific datasets~(such as CUB-200-2011) and encouraging potential on diverse datasets~(such as MS COCO), demonstrating the effectiveness of our proposed ITI~(a paradigm of CMC). It is worth mentioning that we can use a more powerful image generation model to improve the reconstruction performance, but the main aim in this work is to demonstrate the effectiveness of cross modal compression, so improving the submodules of CMC is beyond the scope of this work.

\subsection{Quantitative Results}
% \textbf{JPEG/JPEG2000 Baseline.} 
To evaluate the compression performance of our proposed framework quantitatively, we compare our proposed ITI with the widely used JPEG~\cite{wallace1990overview} and JPEG2000~\cite{marcellin2000overview} standards. 
%We use Python PIL library~\footnote{https://pillow.readthedocs.io/en/stable/} to compress the images with different quality factors. Then
 % We can get a rate-distortion curve by compressing the image with different quality factors, which is used as a baseline for comparison, just like the R-D curve in the traditional image/video compression. Specifically, 
We compress all the images in the testing set with different quality factors, then plot the R-D curve, just like the R-D curve in the traditional image/video compression, as shown in Fig.~\ref{fig:end_to_end}.
Among the four metrics, IS and FID are set level metrics, IPD is an instance level metric, PSNR is a pixel level metric.
Some conclusions can be drawn from the quantitative comparison results:
\\
\indent\textit{ITI surpasses the JPEG baseline and is comparable with JPEG2000 baseline in set level and instance level.} As illustrated in Fig.~\ref{fig:end_to_end} (a), (e), our proposed ITI achieves a similar IS score with a rather lower bitrate when compared with JPEG baseline on both MS COCO and CUB-200-2011 datasets. Also, our proposed method is comparable with JPEG2000 if the trend of the curve is taken into consideration.
% On MS COCO dataset, our CMC model can achieve an IS score of 9.41 with an average bitrate of 25.41 bytes, which outperforms the JPEG baseline curve. On CUB-200-2011, although the IS score of our CMC model is only comparable with the lowest two bitrate points, our bit rate is rather lower than the JPEG baseline curve, hence our CMC model surpassed the JPEG baseline. 
From Fig.~\ref{fig:end_to_end} (b) and (f), better FID scores are obtained by our proposed method than both JPEG and JPEG2000 baselines. IS and FID are set level metrics, so our ITI model shows better performance than the JPEG baseline and comparable performance with JPEG2000 when they are evaluated with set level metrics. IPD, as defined in Eq.~\ref{eq:IPD}, is an instance level metric, measuring the perceptual distance between two samples. As illustrated in Fig.~\ref{fig:end_to_end} (c) and (g), our ITI model can even achieve better comparison results than both JPEG and JPEG2000 baselines. 
% On MS COCO dataset, only two bitrate points among the total six bitrate points of the JPEG baseline show better IPD than our CMC, not to mention that our CMC is with a rather lower bitrate. On CUB-200-2011 dataset, our CMC surpasses all the six bitrate points of the JEPG baseline. 
So our ITI proposed model~(a paradigm of CMC)~shows better performance than JPEG and comparable performance with JPEG2000 when evaluated with instance level perceptual metrics. 
% Overall, 
% \\
% \indent\textit{Our proposed ITI is comparable with JPEG baseline in pixel level.} As illustrated in Fig.~\ref{fig:end_to_end} (d) and (h), 
% our ITI model can achieve only 27.91/27.91 dB on MS COCO/CUB-200-2011 dataset, which is lower than all the bit rate points of JPEG baseline. 
% It is impossible to align the bitrate of the JPEG baseline into our results, we can only compare our results with JPEG baseline. 
% with the help of JPEG trendline~\footnote{we fit the data with cubic polynomial.}, we can see that our ITI model is comparable with the JPEG baseline. So our ITI model is comparable with the JPEG baseline in pixel level when evaluated with the traditional PSNR metric.

\indent\textit{Our proposed ITI can achieve ultrahigh compression ratio with set/instance level reconstruction.} As illustrated in Fig.~\ref{fig:end_to_end}, the bitrate is as low as 26.28/36.69 bytes on MS COCO/CUB-200-2011 dataset, which is rather lower than that in JPEG and JPEG2000 baselines. Our ITI model's set/instance level reconstruction performance~(IS, FID, and IPD) is comparable with the JPEG and JPEG2000 baselines at the time of low bit rate. but our compression ratio is ultrahigh due to the low bitrate. So our ITI model~(a paradigm of CMC) has the potential for the applications which require an ultrahigh compression ratio and only instance level or set level reconstruction, such as image/video data transmission for machine analysis with low bandwidth.

% \section{Ablation Study}\label{sec:ablation_study}

\section{Conclusion and Future Works}\label{sec:conclusion}
In this paper, we proposed~\textit{Cross Model Compression~(CMC)}, a novel compression framework towards human-comprehensible semantic compression. Compared with prior related frameworks, including traditional compression, ultimate feature compression, and intermediate feature compression, CMC is human-comprehensible and can be directly used in semantic monitoring. In particular, a paradigm of CMC,~\textit{Image-Text-Image~(ITI)} is implemented for image compression. Qualitative and quantitative results showed that ITI can achieve an ultrahigh compression ratio and outperformed the JPEG baseline in set level and instance level. Our experimental results demonstrated that CMC has the potential for the applications which require an ultrahigh compression ratio and only set level or instance level reconstruction is required, such as image/video transmission for machine analysis with low bandwidth. 
% In the future, we will continue exploring CMC for video compression, which is more meaningful and necessary for practical applications.
Although encouraging results have been achieved, there are still several open problems related to CMC that we will continue exporing:
\begin{enumerate}
  \item \textit{End-to-end Optimization.} In this work, CMC encoder, and CMC decoder are optimized subsequently, which will result in suboptimal performance. End-to-end optimization can improve the performance further.
  \item \textit{CMC for Video Compression.} Compared with image data, video data are rather more redundant because of their time sampling. The video compression paradigm with higher compression performance is urgent due to the explosive growth of the video data on the Internet. CMC for video compression is more meaningful and necessary for the practical applications, but training deep models on video data is more difficult than that on image data.
  \item \textit{Scalable Compression.} In most image/video transmission scenarios, the network is fluctuant. Therefore rate control is necessary for the compression paradigm. Scalable CMC will make our paradigm more practical and bridge the gap between our experiments and practical applications.
 \item \textit{Semantic-based Metrics.} Although some semantic-based metrics have been proposed, the semantic-based compression frameworks have not been well evaluated so far. Better semantic-based metrics, especially differentiable metrics, are required for optimizing the compression model.
\end{enumerate}
% We will continue exporing these open problems to make the CMC more practical and efficient.

\begin{acks}
% The authors would like to thank Dr. Yuhua Li for providing the matlab code of the \textit{BEPS} method.
% The authors would also like to thank the anonymous referees for their valuable comments and helpful suggestions.
This work was supported in part by the~{National Natural Science Foundation of China} under Grant No.~\grantnum{}{62025101} and No.~\grantnum{}{62088102},~{National Postdoctoral Program for Innovative Talents, China} under Grant No.~\grantnum{}{BX2021009},~{China Postdoctoral Science Foundation}~under Grant No.~\grantnum{}{2020M680238},~and High-performance Computing Platform of Peking University, which are gratefully acknowledged.
% This work is supported by the \grantsponsor{GS501100001809}{National Natural Science Foundation of China}{https://doi.org/10.13039/501100001809} under Grant No.:˜\grantnum{GS501100001809}{61273304} and˜\grantnum[http://www.nnsf.cn/youngscientists]{GS501100001809}{Young Scientists’ Support Program}.
\end{acks}
\bibliographystyle{ACM-Reference-Format}
\bibliography{sample-base}
\end{document}